# Title: Rigid Particle-Induced Anomalous Softening in Elastomer Composites

**Authors:** Yonghyun Cho[1]†, Hyeongho Min[2,3]†*, Hongdeok Kim[1]†, Dongeun Lee[1], Minjune Kim[1], Siyeon Jang[2,4], Ali Amini[2], Zitang Wei[3], Hosub Lim[5], Hwan-ho Cho[6], Bo-yong Park[7,8], Sang Won Lee[1], Joonmyung Choi[1]*, Metin Sitti[2,9]*, Sangyul Baik[1]*

## ABSTRACT

Particle-filled soft composites generally stiffen as particle loading increases, imposing a trade-off between functionality and compliance. Here, we show that this trade-off can be overcome by designing particles to actively direct gelation rather than act as passive fillers. In a sol-rich elastomer matrix, sol-phase kinetics redirect network formation towards particle-mediated assembly, yielding anomalous softening despite increasing particle content. This response is governed by topological reconfiguration of the load-bearing network during gelation rather than by filler loading alone. A structure-based model captures the composition-dependent onset of this behaviour, and the same response is reproduced across distinct particle systems. The resulting materials combine functionality with softness, energy dissipation, and recoverability under large deformation. These findings establish topology-guided gelation as a general strategy for engineering compliant functional composites for soft robotic actuation, conformal interfaces, and other systems that require both deformability and embedded material functionality.

## MAIN

Adding rigid particles to soft composites makes them stiffer (1, 2). Engineers have attempted to design soft composites to overcome the limitations of single materials by simultaneously achieving a lightweight structure (3), softness (4), and various functionalities (electrical (5), magnetic (6, 7), thermal (8), optical (9), etc.); however, this simple paradigm has been the fundamental premise guiding composite research for several decades. The mechanism by which particles reinforce the matrix network—particularly the stiffness of gel-based networks—has been interpreted from a static perspective focused on particle–matrix interactions (10, 11). As a result, research on composite materials has largely operated within the constraint of a functionality–compliance trade-off among the functions of the main components (12, 13). However, the essence of soft composites lies not in simple network reinforcement but in the continuous, nonlinear, and dynamic interactions among its three constituents, namely, the gel network, rigid particles, and sol, whereas the mobile polymer fraction is not incorporated into the permanent gel network (14, 15). In this context, while the gel is responsible for structural support and elastic recovery (16, 17), the sol has primarily been regarded as an energy dissipation pathway; however, its significance lies in its fluidity, which mitigates structural constraints and determines the system's deformability (15). This perspective offers a crucial clue for transcending existing frameworks.

Nevertheless, it remains unclear whether hard particles can be used to make composites softer. Mechanical compliance is typically associated with low solid content or loosely crosslinked networks (18, 19). In heterogeneous composites, however, the mechanical response depends on both the composition and the connectivity of stress-transmission pathways (20–23). In other words, what matters is not only how much of each component is present but also how these components are connected. Although the role of network connectivity in determining macroscopic mechanical properties is well recognized, previous studies have focused primarily on the structure and mechanical response of already formed networks. Consequently, the connectivity established during network formation itself has received comparatively limited attention as a design variable. However, the fluidity of the sol offers a different perspective. During network formation, this mobility allows the constituent phases to move and rearrange, with potential consequences for the connectivity and stress-transmission pathways that ultimately emerge (24). The key therefore lies in the mobility and rearrangement potential of the components that determine structural connectivity during network formation.

### Increased particle loading induces anomalous softening

Starting from the possibility that particles can influence a matrix while it is still forming, we designed a soft composite in which gelation delay preserves network mobility before fixation (Fig. 1a). This mobile state provides a window during which the forming gel network can be spatially reorganized rather than being immediately fixed throughout the bulk. We propose that particle interfaces bias this organization towards particle-centred gel domains such that increasing particle loading changes not only the number of rigid inclusions but also how the load-bearing matrix is formed. In this framework, particles are therefore considered not simply reinforcements of a preexisting network but spatial regulators of the forming gel architecture.

The resulting mechanical response is opposite that of conventional particle reinforcement. Under the same spherical load, a composite with low particle loading showed little deformation, whereas a composite with higher particle loading showed substantially more deformation (Fig. 1b), directly demonstrating increased compliance despite the greater content of rigid particles. This inversion changes the conventional relationship between particle-derived functionality and mechanical stiffness: although increasing functional-particle loading is generally accompanied by a higher modulus, the present system exhibits increasing particle-derived functionality while the mechanical resistance is reduced (31–33) (Fig. 1c). This anomalous softening therefore creates a composition space in which higher functional-particle loading can coexist with mechanical compliance.

We hypothesize that this counterintuitive response originates from a particle loading-dependent redistribution of the load-bearing gel topology (Fig. 1d). At lower particle loading, a larger fraction of the gel can participate in extended connections across the interparticle space. As the particle loading increases, the gel is increasingly organized into particle-centred domains, leaving fewer continuous connections available for long-range stress transmission. The resulting shift from a bridge-rich connected network towards a more island-dominant architecture provides a conceptual explanation for how adding rigid particles can reduce, rather than increase, the stiffness of the composite. We next examine the composition range over which this anomalous response emerges, how the gelation delay redirects network formation towards particle interfaces and how the resulting topology governs the elastic modulus.

**Anomalous softening accompanies network reorganization**

Sol enrichment reverses the mechanical response to particle loading, converting conventional reinforcement into anomalous softening. We tuned the Ecoflex-30 matrix towards a sol-rich condition by increasing the content of polyethyleneimine ethoxylated (PEIE); scanning electron microscopy (SEM) imaging and measurements of the extractable fraction and swelling ratio revealed progressive sol enrichment with increasing PEIE content (Extended Data Fig. 1 and Supplementary Fig. 1). At 0.3 wt% PEIE, the stress level increased with increasing particle fraction, which is consistent with conventional particle reinforcement (Fig. 2a, i). At 1.2 wt% PEIE, the response became nonmonotonic: the mechanical response initially increased with increasing particle loading at low particle fractions, then decreased over an intermediate loading range and subsequently increased at higher particle fractions (Fig. 2a, ii). At 1.5 wt% PEIE, the suppressed stress response extended across a broader accessible particle loading range (Fig. 2a, iii).

To further investigate this inversion, we examined the corresponding Young's modulus profiles (Fig. 2b). At 1.5 wt% PEIE, for example, the Young's modulus decreased from approximately 25 kPa without particles to approximately 9 kPa at 60 wt% particle loading, despite the increasing content of rigid particles. To compare the composition range over which this response occurs, we define the softening onset, $P_s$, as the particle fraction at which the modulus begins to decrease. At higher particle fractions, some compositions subsequently exhibit modulus recovery, with $P_g$ denoting the point at which the modulus begins to increase again after reaching a minimum. As the PEIE content increased, $P_s$ shifted towards lower particle fractions (red arrow in Fig. 2b), whereas re-stiffening, when observed, was displaced towards higher particle loading (blue arrow in Fig. 2b). The sol-rich state of the matrix therefore controls not only the modulus of the composite but also the particle loading range over which rigid particle addition yields anomalous softening.

Anomalous softening is accompanied by pronounced reorganization of the internal gel network without a comparable change in the gel fraction. X-ray microtomography (XRM) revealed distinct gel–sol morphologies across the particle loading range (Fig. 2c). At low particle loading, the gel phase formed broad, interconnected regions spanning the matrix. Within the softening regime, these regions became increasingly fragmented into smaller and more spatially localized gel domains separated by sol-rich regions. At higher particle loading, the connected gel features re-emerged, indicating that structural evolution is not a simple monotonic disruption of the network. Importantly, the amount of cured gel did not follow the pronounced changes in either the morphology or modulus. At 1.5 wt% PEIE, the gel fraction remained within approximately 39–43% as the particle loading increased from 10 to 60 wt%, whereas the Young's modulus decreased from approximately 22 to 9 kPa over the same loading range (Fig. 2b and d). Therefore, the modulus reduction cannot be attributed simply to the formation of less gel. Instead, the contrast between the nearly conserved gel fraction and the strongly evolving morphology on XRM reveals that the spatial organization and connectivity of the gel phase are relevant structural variables. Loading-induced mechanical inversion was also reproduced in chemically distinct NdFeB- and alumina-filled composites, indicating that anomalous softening is not specific to CIPs (Supplementary Figs. 2 and 3). These observations associate anomalous softening with how an approximately conserved gel phase is spatially organized within the composite rather than with the amount of gel alone.

The softened state remains strongly dissipative under large deformation rather than behaving as a simply degraded network. During cyclic loading to 150% strain, the first loading cycle produced initial mechanical conditioning, after which the stress–strain hysteresis stabilized. From the second cycle onwards, the successive loading–unloading curves showed little change over repeated deformation (Fig. 2e; the complete cyclic response, including the first cycle, is shown in Supplementary Fig. 4). Therefore, the softened composite reaches a stable cyclic state despite its low modulus. This stability is also maintained under substantially larger deformation: a representative high-particle-fraction sample could be stretched to 800% strain and subsequently recover its macroscopic shape after unloading (Fig. 2f). The coexistence of a reduced modulus, cyclic mechanical stability, and large-strain recoverability distinguishes anomalous softening from a simple loss of structural integrity. This raises the mechanistic question of how increasing the particle content can reduce stiffness while the gel fraction remains nearly unchanged and the mechanical response remains stable.

**Gelation delay redirects network formation towards particle interfaces**

Gelation delay redirects the spatial organization of the forming gel network towards particle interfaces. Introducing PEIE delayed gelation and extended the period over which the matrix remained mobile before network fixation (Fig. 3a). To directly examine the spatial organization of the gel under these conditions, we used a composition containing 1.5 wt% PEIE and 70 wt% particles. At this composition, particle-associated gel domains are formed but remain insufficiently interconnected to establish a system-spanning network, resulting in a macroscopic gum-like state. The surrounding extractable phase was removed through solvent washing, allowing the particle-associated gel to be isolated from the nonpercolated matrix structure. SEM subsequently revealed a distinct gel-rich layer retained around individual particle surfaces (Fig. 3b). The gum-like state therefore provides direct structural evidence that the gel can become locally concentrated around particles without being incorporated into a continuous bulk network.

This particle-proximal accumulation can be described by kinetic competition between bulk association and particle-interface capture during the prolonged mobile-sol window. Mobile network-forming species can either encounter one another and associate in the bulk or encounter a particle surface and become captured at the interface (25) (Fig. 3c). In the coarse-grained kinetic description, this bulk association was represented as an effective second-order pathway because it requires encounters between mobile species, whereas particle-interface capture was represented as an effective first-order pathway governed by encounters between individual species and the particle surface (Supplementary Text S1). Extending the mobile-sol window therefore provides repeated opportunities for particle–surface encounters before the bulk association fixes the surrounding network.

Subsequently, molecular dynamics (MD) simulations were performed to examine how this kinetic competition translates into particle-associated molecular retention and spatial redistribution. The distance of the molecular mass centre from the particle depended strongly on the molecular weight (Fig. 3d). Higher-molecular-weight species remained localized closer to the particle, whereas lower-molecular-weight species underwent larger positional fluctuations and sampled configurations whose average positions extended farther towards the bulk, indicating. Consistently, the calculated particle–species interaction energy increased with increasing molecular weight (Supplementary Fig. 5), supporting the stronger retention of larger network-forming species at the particle interface. The simulated gel fraction trajectories reached a plateau by $10^8 t^*$, confirming that the subsequent structural comparisons were performed after the simulated gelation process had reached a stable endpoint (Supplementary Fig. 6). MD simulations of network formation further revealed that these molecular-scale interactions resulted in the progressive concentration of these network-forming species around particle surfaces rather than a spatially uniform distribution (Fig. 3e). This redistribution was quantified using radial gel fraction profiles extracted from the MD-generated structures. The modelled reaction rate decreased from P0 (100%) to P1 (50%) and P2 (1%), representing progressively slower network fixation. As the reaction rate decreased from P0 to P2, the gel fraction became progressively enriched near the particle relative to regions farther into the bulk (Fig. 3f). The MD results therefore link prolonged molecular mobility and particle-associated retention to the emergence of particle-centred gel-rich domains.

The MD-derived single-cell analysis further revealed that particle-centred gel accumulation does not necessarily translate into a continuous load-bearing state. Quantitative deformation analysis revealed that compared with the corresponding particle-free gel cell, a particle-containing cell exhibited lower deformation resistance (Fig. 3g). Thus, although the gel is locally concentrated around the particle, this configuration does not behave as a stronger load-bearing gel cell. Instead, the reduced deformation resistance is consistent with the incomplete integration of the particle-associated gel into the continuous stress-transmitting network. This distinction between local gel accumulation and network-level load bearing motivates the examination of how increasing particle number reorganizes system-spanning gel connectivity.

**Topology reorganization governs the elastic modulus**

Increasing particle loading redistributes the gel phase from bulk-spanning regions towards particle-centred domains. At low particle fractions, the gel formed broadly throughout the interparticle space, resulting in an extensively connected morphology, whereas increasing the number of particles resulted in the gel becoming progressively concentrated around individual particle interfaces (Fig. 4a). This redistribution was also observed in the distance-dependent gel fraction profiles obtained for particle arrays containing 2 × 2, 3 × 3, and 4 × 4 particles (Fig. 4b). In sparse arrays, a larger fraction of the gel remained distributed away from the particle surfaces, whereas increasing the particle number progressively enriched the particle-proximal regions and depleted the gel from the intervening bulk. Thus, increasing particle loading redistributes the same gel resource from bulk-spanning regions towards interface-localized domains.

This redistribution partitions the cured gel into structurally distinct components with different roles in system-spanning load transfer. XRM and SEM revealed particle-centred gel domains, interconnecting bridges, dangling structures, and disconnected regions (Fig. 4c). On the basis of these experimentally observed morphologies, we decomposed the gel structure into particle-centred islands, interconnecting bridges, dangling or disconnected domains, and the largest connected component (LCC), which formed the system-spanning backbone. Whereas isolated or dangling components do not directly participate in system-spanning load transfer, connected bridges within the LCC provide pathways through which stress can be transmitted across the material (17, 26). This structural decomposition therefore provides a basis for distinguishing the amount of gel from the fraction of gel incorporated into the load-bearing connected structure.

Within the anomalous softening regime, increasing particle loading progressively reduces the connectivity of this load-bearing network. Skeleton-based analysis of the XRM structures was used to quantify the corresponding topology descriptors, including island- and bridge-associated features (Extended Data Fig. 2a). Specifically, $z_c$ denotes the number of load-bearing bridges within the connected backbone, whereas $r$ represents the spatial span of the connected gel network. $A_e/L_e$ provides a geometric proxy for the effective stiffness of the remaining bridges because the stiffness of a bridge scales as $E_{\mathrm{gel}}A_e/L_e$. As the particle fraction increased, island-associated descriptors increased, whereas bridge-associated descriptors decreased, accompanied by systematic changes in the geometry and connectivity of the remaining bridges (Fig. 4d and Supplementary Fig. 7). The corresponding topology maps and backbone graphs, reconstructed directly from experimentally measured XRM structures, visualize the same progression across representative composition states, showing increasing fragmentation of the system-spanning network as the particle loading increases (Fig. 4e). Importantly, this topology trajectory was PEIE-dependent and therefore cannot be explained by increasing particle crowding alone. At matched CIP loadings of 20, 40, and 60 wt%, the evolution of the gel–sol morphology of the PEIE-containing composites differed markedly from that of the PEIE-free controls, with stronger redistribution of the gel phase away from extended bulk-spanning pathways towards spatially localized domains (Supplementary Fig. 8). The structural evolution therefore reflects a transition from a bridge-rich connected morphology towards an increasingly island-dominant topology with fewer effective load-bearing connections.

The loss of load-bearing connectivity accounts for the decrease in the elastic modulus during anomalous softening. In a heterogeneous network containing islands and bridges, macroscopic elasticity is governed primarily by the connected structures capable of transmitting stress across the material (23, 27). Within this backbone-centred framework, the modulus depends on both the number of elastically active bridges belonging to the LCC and the effective stiffness of those bridges, which is determined by their geometry (Supplementary Text S2 and Extended Data Fig. 2b). In the topology-reconfigured composites, increasing the particle loading progressively softened the material as the load-bearing network became less effective. In contrast, the PEIE-free conventional composites stiffened with increasing particle loading, which is consistent with the formation of a more mechanically effective load-bearing network (Supplementary Fig. 9). This comparison indicates that anomalous softening arises from a distinct evolution of the load-bearing topology during network formation rather than from particle loading alone. As the particle loading increases within the softening regime, the number of effective bridges decreases, and the remaining connections become less mechanically effective. A structure-based modulus scaling constructed from these experimentally derived descriptors reproduced the observed decrease in the normalized modulus with increasing particle fraction (Fig. 4f). The anomalous softening can therefore be accounted for by the particle-induced redistribution of the gel away from the bulk-spanning load-bearing pathways and towards locally concentrated domains, reducing the effective connectivity of the elastic backbone despite the nearly unchanged gel resource.

**Topology-reconfigured softening expands accessible functional-particle loading**

Topology-reconfigured softening extends across chemically and functionally distinct particle systems, enabling higher particle loading with a substantially reduced mechanical penalty. We compared conventional and topology-reconfigured composites containing alumina, carbonyl iron particles (CIPs), copper (Cu), neodymium–iron–boron (NdFeB), and silver (Ag) using a functional gain-to-modulus index that relates the increase in particle loading to the accompanying change in the Young's modulus. Because these particles have different physical functions, particle loading was used as a common compositional proxy for the amount of particle-derived functionality available to the composite rather than as a direct measure of functional output. Conventional composites showed limited index values because increasing particle content was accompanied by an increase in the modulus. In contrast, topology-reconfigured composites exhibited substantially larger values as this stiffness penalty was reduced or reversed (Fig. 5a). The inverted loading–modulus relationship therefore extends beyond single-particle chemistry and expands the range of functional-particle loading that can be incorporated while retaining mechanical compliance.

This expanded loading range directly benefits magnetic actuation, for which both the magnetic-particle content and mechanical resistance constrain deformation. A topology-reconfigured composite containing 60 wt% NdFeB underwent pronounced deformation under a 5-mT magnetic field, whereas a conventional composite containing 20 wt% NdFeB showed little deformation under the same field (Fig. 5b and Supplementary Video 1). In this reconfigured composite under high loading, this low-field deformation was converted into crawling locomotion under sequential magnetic-field inputs. Thus, increasing the amount of magnetic material while simultaneously reducing elastic resistance allows the applied magnetic field to produce substantially greater macroscopic deformation, illustrating how anomalous softening can convert increased functional-particle loading into an accessible actuation response.

The same mechanical advantage is evident for materials whose functions depend on maintaining intimate interfacial contact during deformation. In alumina-filled composites, compared with the conventional 20 wt% composite, the topology-reconfigured material containing 60 wt% particles conformed to the stepped chip topography, and interfacial air gaps were reduced (Fig. 5c). This closer contact produced a warmer and more spatially uniform thermal response, showing how reduced stiffness enables a highly particle-loaded thermal material to better accommodate a structured interface.

Additionally, in the CIP-filled shielding system, the topology-reconfigured 60 wt% composite maintained conformal contact with a foldable device during bending and suppressed the measured electric field signal, whereas the conventional 20 wt% composite partially delaminated and exhibited larger signal fluctuations (Fig. 5d). Thus, across magnetic actuation, thermal interfacing, and electromagnetic shielding applications, topology-reconfigured softening makes this higher functional-particle loading a mechanically usable property by reducing the stiffness constraints that otherwise limit deformation and interfacial conformity.

## Discussion

The anomalous softening observed here changes the conventional understanding of how rigid particles determine the mechanics of soft composites. Classic inclusion-based descriptions provide a framework for considering particles as mechanically distinct inclusions embedded within an established matrix (28, 29). Our results show that the mechanical consequences of particle addition are not predetermined when the matrix remains structurally mobile during network formation. Under gelation-delayed conditions, increasing particle loading can alter how the forming gel phase is spatially organized rather than simply adding reinforcement to a fixed network. Importantly, the pronounced decrease in the modulus occurs without a comparable reduction in the gel fraction, indicating that the amount of gel alone cannot account for the mechanical response. Instead, the elastic response depends on how the approximately conserved gel phase is partitioned between locally concentrated domains and system-spanning connections. At higher particle loading, the re-emergence of connected structures can coincide with modulus recovery, further emphasizing that the particle content and stiffness are linked through the evolving organization of the matrix rather than through the particle fraction alone.

This behaviour can be attributed to competition during network formation. Gelation delay extends the period during which network-forming species remain mobile, increasing the opportunity for repeated interactions with particle interfaces before the surrounding network becomes fixed. Under these conditions, the capture and retention behaviour at the particle–interface boundary bias the spatial distribution of the forming gel towards particle-centred domains. As the particle number increases, a greater portion of the approximately conserved gel phase is localized around the particles, leaving fewer effective connections available to form a system-spanning load-bearing backbone within the softening regime. Fillers and additives have previously been shown to alter polymer curing kinetics and, in silicone systems, to interfere with catalytic curing reactions (30, 34–36). The distinction here is not simply that particles or additives affect curing. Rather, a prolonged mobile state converts particle–matrix interactions during curing into a spatial redistribution of network formation that is sufficiently pronounced to reorganize load-bearing connectivity and reverse the usual loading–stiffness relationship over an accessible composition range. The network topology during this formation process therefore emerges as a mechanical design variable in addition to the material composition.

The consequence of this topology-based control extends beyond the mechanical anomaly itself. Similar loading–modulus inversion across chemically and functionally distinct particle systems indicates that the response is not restricted to single-particle chemistry. More importantly, reducing the stiffness penalty associated with increasing particle content makes higher functional-particle loading mechanically accessible, as demonstrated by magnetic actuation, thermal interfacing, and electromagnetic shielding experiments. In these systems, the benefit does not arise from particle loading alone but from combining increased particle-derived functionality with the compliance needed for deformation or conformal contact. Particle loading can therefore serve not only as a means of increasing functionality but also as a method to control mechanical compliance through its influence on network formation. More broadly, treating the matrix-formation history and the resulting load-bearing topology as design variables in addition to the material composition provides a route to expand the accessible property space of functional soft composites.

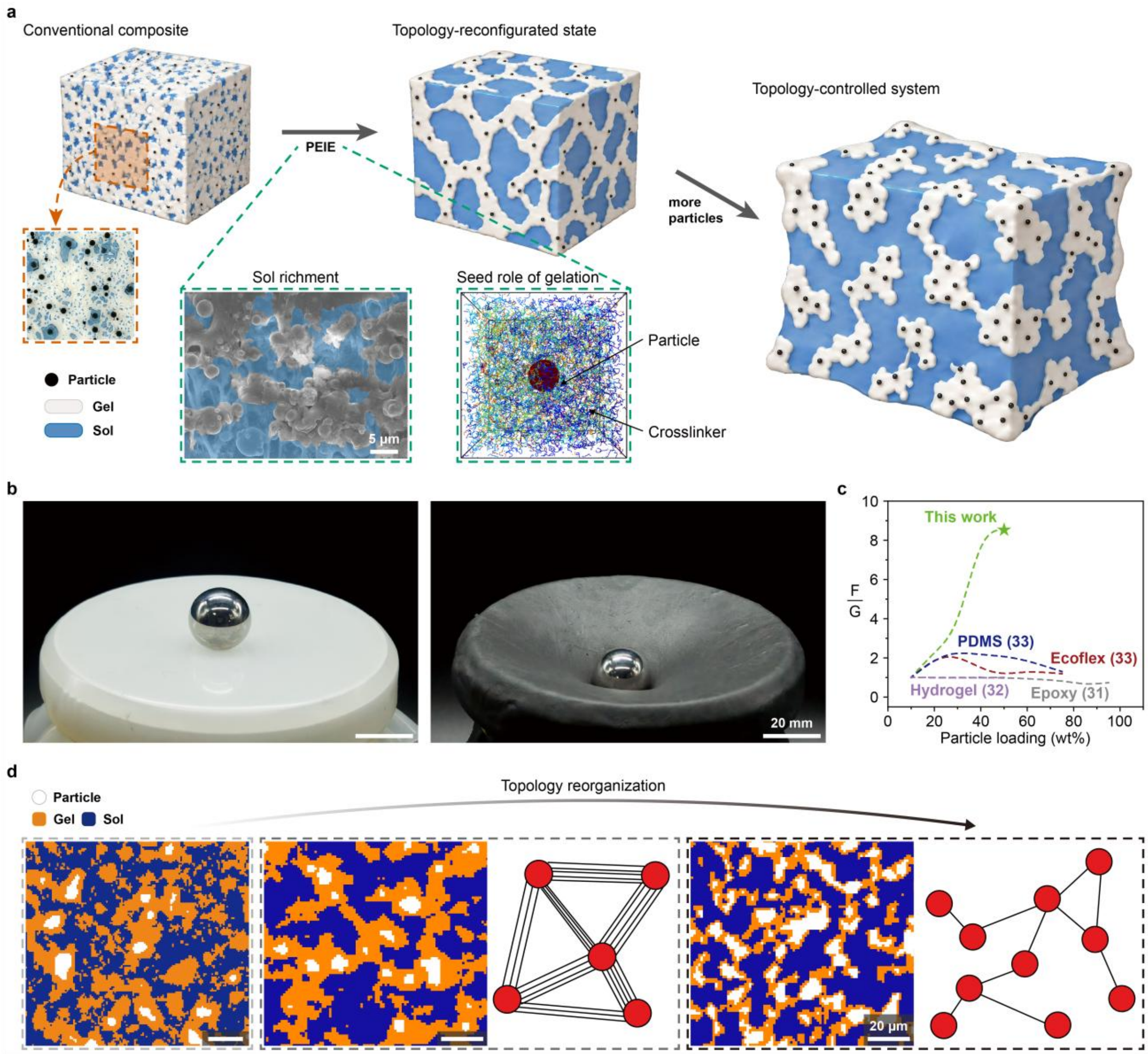

**Fig. 1. Particle-induced topology reorganization enables anomalous softening in elastomer composites. a,** Schematic of the design principle. In conventional particle-filled composites, rigid particles primarily reinforce the surrounding elastomer network. Gelation delay preserves matrix mobility before network fixation, providing a window in which particle interfaces can bias the spatial organization of the forming gel towards particle-centred domains. Increasing particle loading can therefore alter not only the number of rigid inclusions but also the organization of the resulting gel network. The insets show the heterogeneous gel–sol morphology and particle-centred accumulation observed experimentally and in the simulation. **b,** Macroscopic demonstration of anomalous softening. Under the same spherical load, a low-particle-loading composite shows little deformation, whereas a higher-loading composite undergoes pronounced deformation, demonstrating increased compliance despite the greater content of rigid particles. Scale bars, 20 mm. **c,** NdFeB-based comparison of magnetic functionality and mechanical stiffness. Comparison of the functionality-to-modulus ratio (F/G) as a function of particle loading for the material considered in this work and previously reported PDMS-, Ecoflex-, hydrogel-, and epoxy-based composites (31–33). F denotes particle-derived functionality and G denotes the modulus used in the plotted comparison. **d,** Conceptual topology model for the anomalous mechanical response. At lower particle loading, the gel phase forms extended connections across the interparticle space. As the particle loading increases, the gel becomes increasingly organized into particle-centred domains, leaving fewer continuous connections available for long-range stress transmission. The corresponding network representation illustrates the hypothesized progression from a bridge-rich connected network towards a more island-dominant topology with reduced system-spanning connectivity. Orange, gel; blue, sol; white, particles.

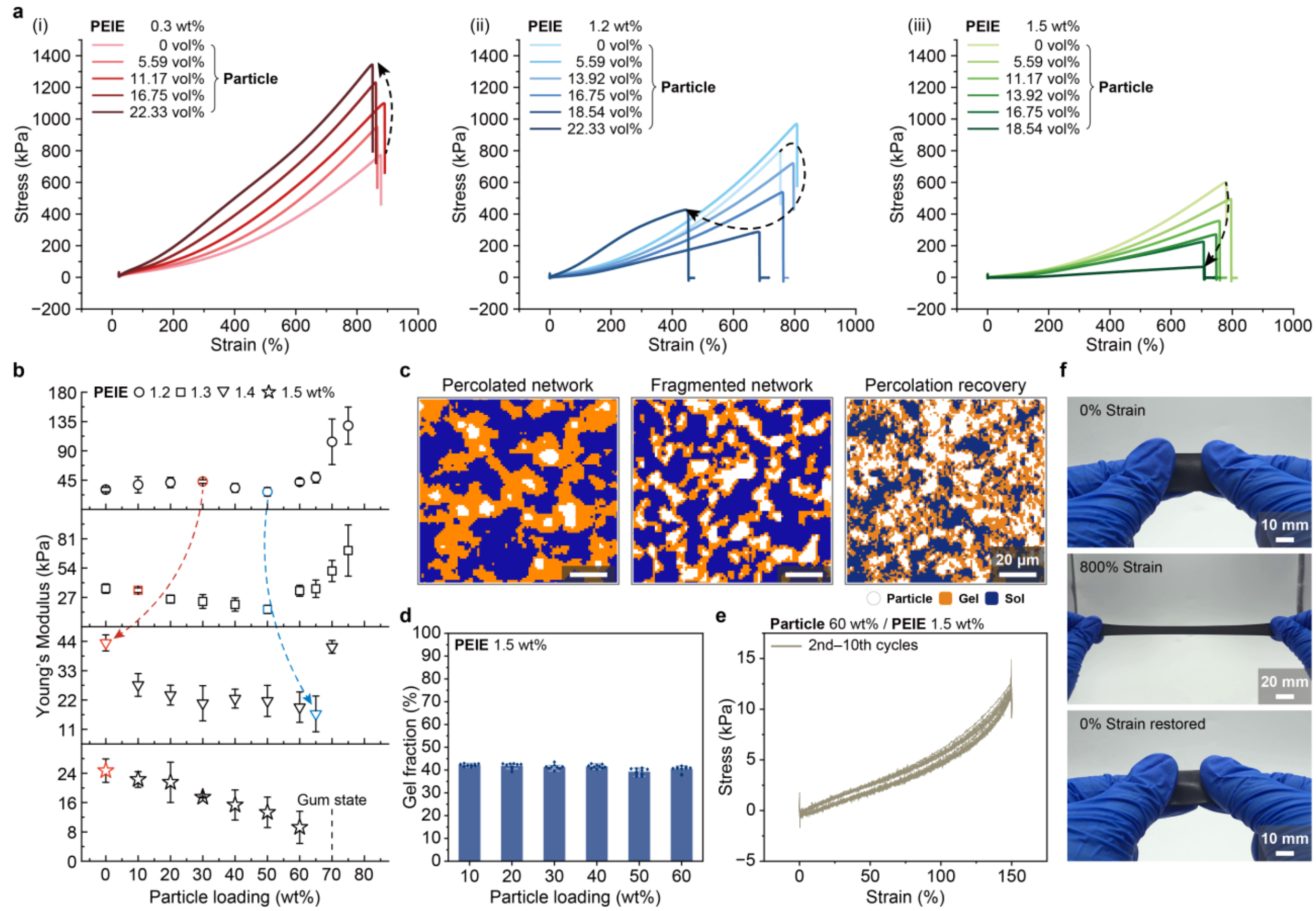

**Fig. 2. Anomalous softening and topology-associated mechanical transition in sol-rich elastomer composites. a,** Representative stress–strain curves at PEIE concentrations of (i) 0.3, (ii) 1.2, and (iii) 1.5 wt%. The particle contents in the legends are expressed as volume fractions (vol%). The curves illustrate conventional reinforcement at 0.3 wt% PEIE and the emergence of loading-dependent softening at higher PEIE concentrations. **b,** Young's modulus as a function of the particle weight fraction. The softening onset, , is defined as the particle fraction at which the modulus begins to decrease, whereas the re-stiffening onset, , denotes the particle fraction at which the modulus begins to increase again after reaching a minimum. Increasing the PEIE content leads to a shift towards lower particle fractions, whereas re-stiffening leads to a shift towards higher particle loading. **c,** X-ray microtomography (XRM) cross-sections showing particle loading-dependent gel–sol organization. At low particle loading, the gel phase forms broad interconnected regions. Within the anomalous-softening regime, the gel becomes fragmented into smaller and more localized domains separated by sol-rich regions. At higher particle loading, connected gel features re-emerge. Orange, gel; blue, sol; white, particles. **d,** Gel fraction as a function of particle loading at 1.5 wt% PEIE, showing that the amount of cured gel remains nearly unchanged despite the pronounced reduction in the Young's modulus. **e,** Stress–strain hysteresis curves for the second to tenth loading–unloading cycles to 150% strain of the composite containing 60 wt% particles and 1.5 wt% PEIE. The complete ten-cycle response, including the first cycle, is shown in Supplementary Fig. 4. **f,** Representative photographs of a high-particle-fraction composite stretched from 0% (top) to 800% strain (middle) and after unloading (bottom), showing large-strain deformability and subsequent macroscopic shape recovery. Scale bars, 10 mm (top/bottom) and 20 mm (middle).

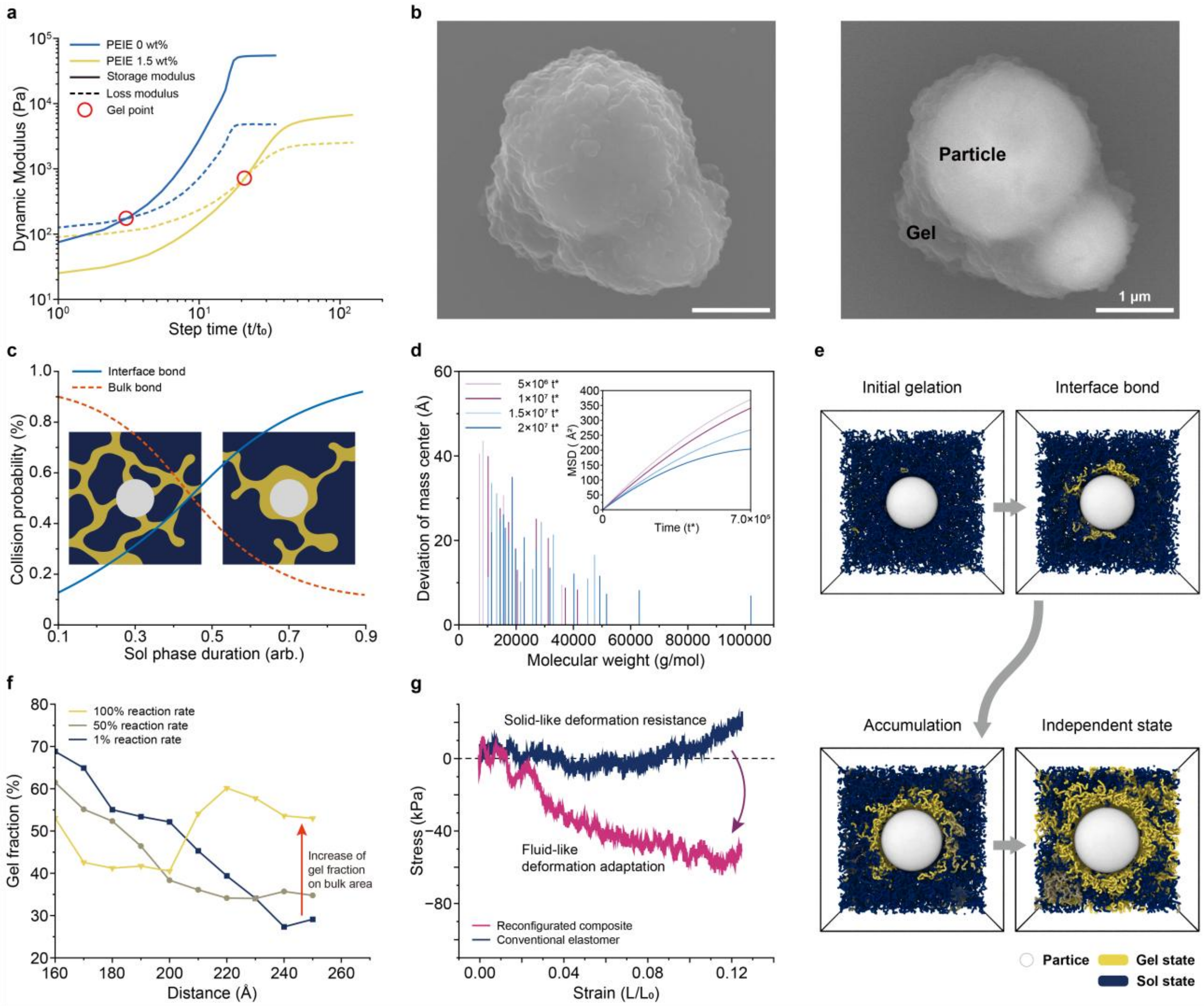

**Fig. 3. Gelation delay redirects network formation towards particle interfaces. a,** Time-resolved dynamic modulus during the curing of matrices without PEIE and with 1.5 wt% PEIE. The addition of PEIE delays the gel point and extends the period over which the matrix remains mobile before network fixation. The dashed circles indicate the gel points. **b,** Scanning electron microscopy (SEM) image of particles isolated from a gum-like composite containing 1.5 wt% PEIE and 70 wt% particles after solvent washing. At this composition, particle-associated gel domains are formed but remain insufficiently interconnected to establish a system-spanning network (left). Removal of the surrounding extractable phase revealed a distinct gel-rich layer retained around individual particle surfaces (right). **c,** Coarse-grained kinetic description of the competing pathways for mobile network-forming species during the extended mobile-sol window. The bulk association is represented as an effective second-order pathway requiring encounters between mobile species, whereas particle-interface capture is represented as an effective first-order pathway governed by encounters between individual species and the particle surface. **d,** Molecular dynamics (MD)-derived deviation of the molecular mass centre relative to the particle as a function of molecular weight. Higher-molecular-weight species remain localized closer to the particle with smaller deviations, whereas lower-molecular-weight species undergo larger positional excursions and occupy average positions farther towards the bulk. **e,** MD simulation snapshots showing the spatial redistribution of network-forming species during network formation. With slower network fixation, species become increasingly concentrated around the particle interface rather than remaining uniformly distributed throughout the bulk. **f,** Radial gel fraction profiles extracted from the MD-generated structures under three modelled reaction rate conditions. P0, P1, and P2 correspond to relative reaction rates of 100, 50, and 1%, respectively. Decreasing the reaction rate from P0 to P2 progressively enriched the gel fraction near the particle relative to the surrounding bulk. **g,** Quantitative deformation response obtained from MD-based single-cell deformation analysis for particle-free and particle-containing gel cells. The particle-containing configuration exhibits lower deformation resistance, indicating that particle-centred gel accumulation does not necessarily result in a stronger load-bearing local state.

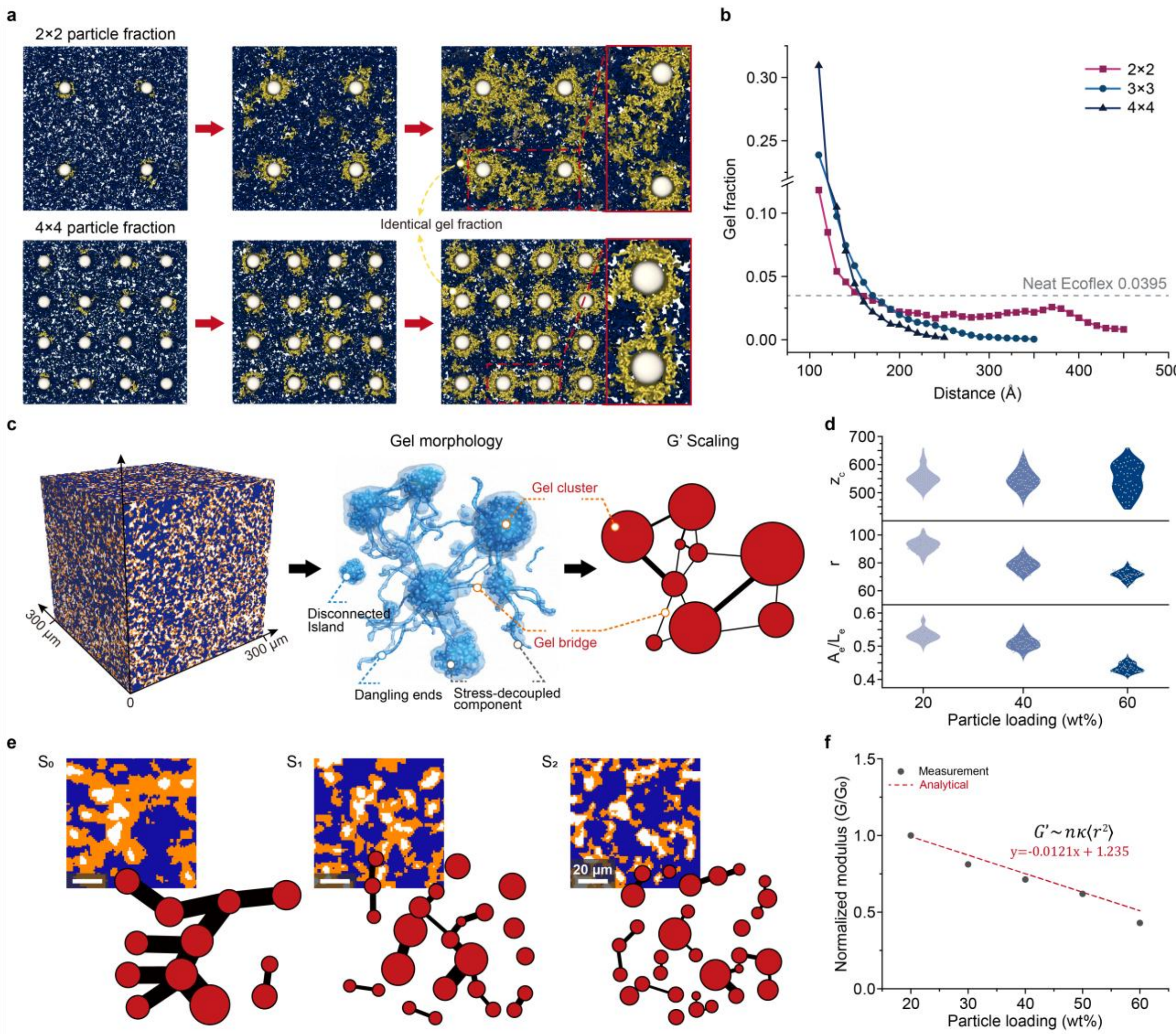

**Fig. 4. Topology reorganization governs the elastic modulus through backbone connectivity.** **a,** Molecular dynamics (MD) simulation snapshots showing particle-number-dependent redistribution of the gel phase during network formation. At lower particle numbers, the gel remains broadly distributed across the interparticle space, whereas increasing the particle number results in the gel being progressively concentrated around the particle interfaces and reduces the amount of gel distributed throughout the intervening bulk. Yellow, gel; navy, sol; white, particles. **b,** Distance-dependent gel fraction profiles extracted from simulated 2 × 2, 3 × 3, and 4 × 4 particle arrays. Increasing the particle number enriches the particle-proximal gel while depleting the gel farther from the particle surface, quantitatively capturing the redistribution shown in (**a**). The dashed line indicates the gel fraction of neat Ecoflex. **c,** Experimental and schematic decomposition of the cured gel morphology into mechanically distinct structural components. Representative XRM image showing particle-centred gel domains together with interconnecting gel bridges. The corresponding schematic display particle-centred islands, interconnecting bridges, dangling or disconnected domains, and the largest connected component (LCC), which provides system-spanning pathways for long-range stress transmission. Scale bar, 10 μm. **d,** Violin plots of the topology descriptors derived from skeleton-based analysis of the XRM structures, including the cluster size ($z_c$), network span (r), and effective bridge stiffness ($A_e/L_e$), as a function of particle

fraction. The collective evolution of these descriptors indicates increasing island dominance and decreasing effectiveness of the interconnecting bridge network with increasing particle loading. **e,** XRM-derived topology maps and corresponding backbone graphs for representative particle loading states S0, S1, and S2, corresponding to 20, 40, and 60 wt% particle loading, respectively. Increasing particle loading results in the progressive fragmentation of the system-spanning backbone and reduces the number of continuous load-transfer pathways. **f,** Normalized Young's modulus as a function of particle loading, comparing experimental measurements with those of the structure-based analytical scaling model. The model incorporates the number and effective stiffness of the load-bearing bridges within the connected backbone and reproduces the experimentally observed decrease in the modulus within the anomalous-softening regime. P0 and P2 denote relative reaction-rate conditions of 100 and 1%, respectively. The decrease in the modulus is reproduced by the backbone-based relation $G' \sim n\kappa(r)$.

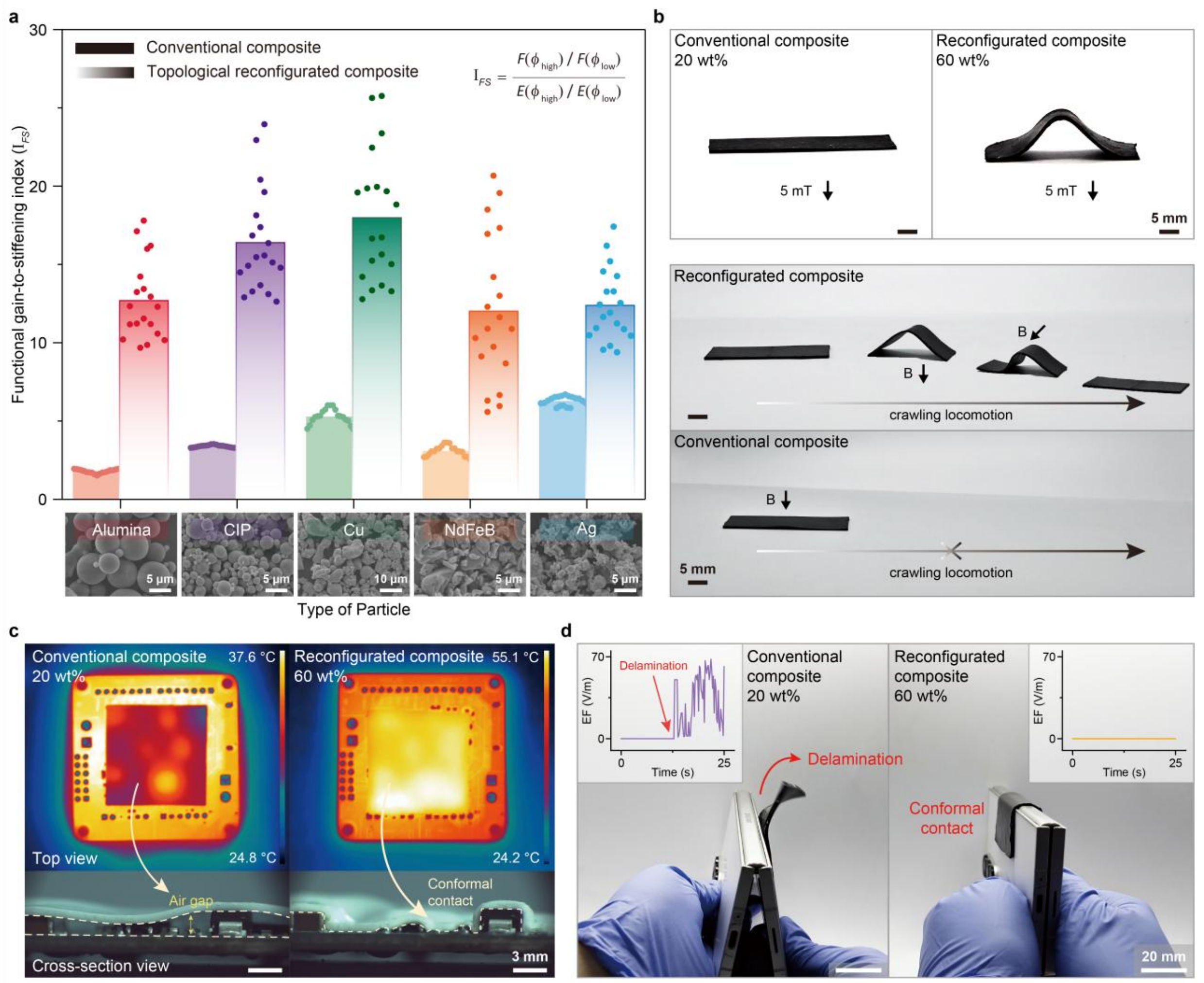

**Fig. 5. Topology-reconfigured softening expands accessible functional-particle loading across diverse composite systems. a,** Functional gain-to-modulus index for conventional and topology-reconfigured composites containing alumina, carbonyl iron particles (CIPs), Cu, NdFeB, and Ag. The index relates the increase in particle loading to the accompanying change in the Young's modulus, with particle loading used as a common compositional proxy for particle-derived functionality rather than as a direct measure of functional output. Larger values indicate a greater increase in accessible functional-particle loading relative to the associated mechanical penalty. The points represent individual measurements, and the bars represent the mean values. Bottom, SEM images of alumina, CIP, Cu, NdFeB, and Ag particles. Scale bars, 5 µm for alumina, CIP, NdFeB, and Ag and 10 µm for Cu. **b,** Magnetic actuation of NdFeB-filled composites. The response of a conventional composite containing 20 wt% NdFeB (left) and a topology-reconfigured composite containing 60 wt% NdFeB (right) under a 5-mT magnetic field was compared. The higher-loading reconfigured composite undergoes pronounced deformation, whereas the conventional composite shows little deformation under the same field. Bottom, crawling locomotion of the topology-reconfigured composite under sequential magnetic field inputs. Scale bars, 5 mm. **c,** Thermal interface of alumina-filled composites on a structured chip surface. The infrared thermographs and corresponding cross-sectional images show a conventional composite containing 20 wt% alumina (left) and a topology-reconfigured composite containing 60 wt% alumina (right). The conventional composite leaves interfacial air gaps, whereas the

topology-reconfigured composite conforms more closely to the stepped surface, producing a warmer and more spatially uniform thermal response. Scale bars, 3 mm. **d,** Electric-field shielding of CIP-filled composites attached to a foldable device. Photographs of a conventional composite containing 20 wt% CIP (left) and a topology-reconfigured composite containing 60 wt% CIP (right) during bending. The conventional composite partially delaminates, whereas the topology-reconfigured composite maintains conformal contact with the device. The insets show the corresponding electric field intensity measured during folding, with the reconfigured composite suppressing the measured signal and reducing fluctuations associated with loss of contact. Scale bars, 20 mm.

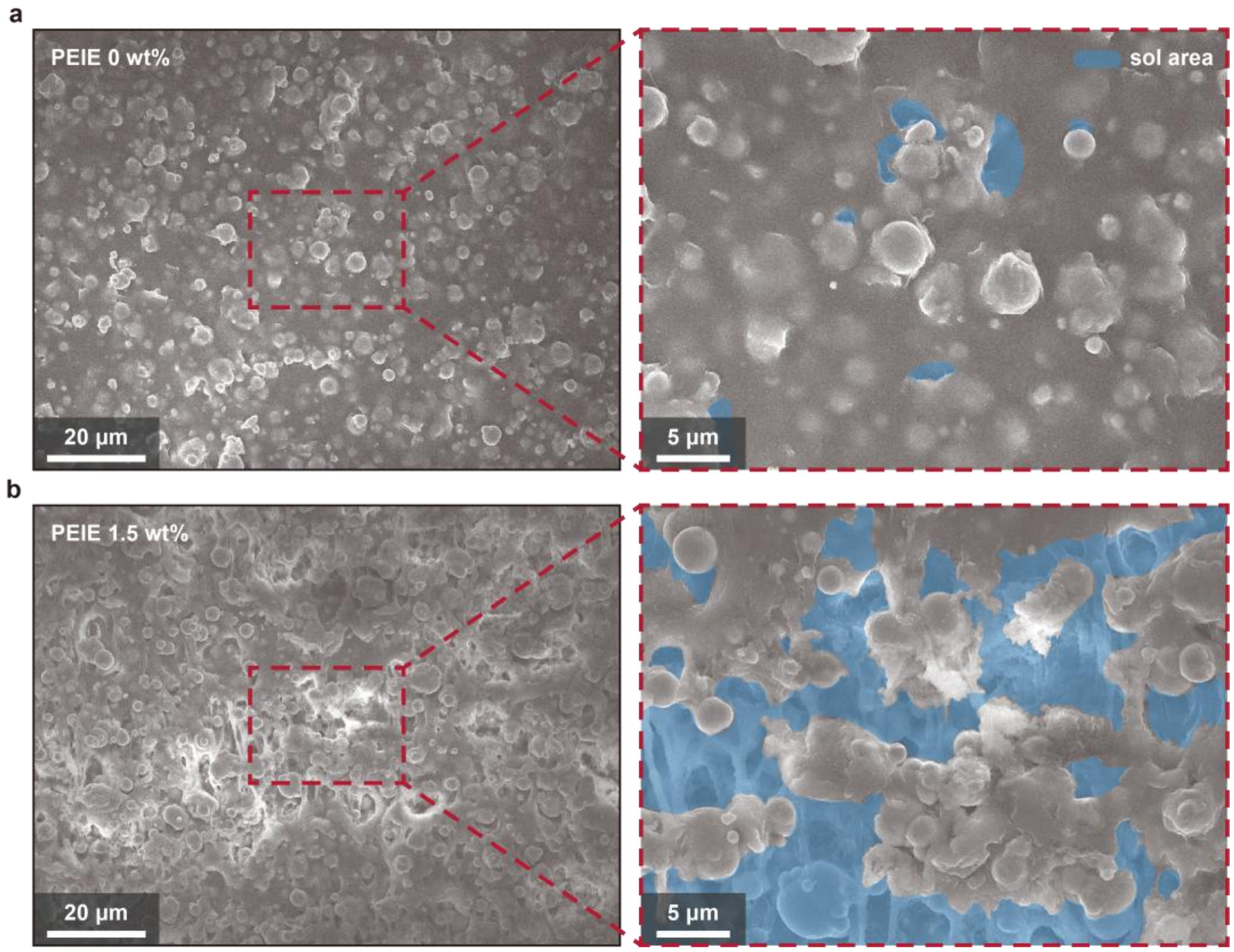

**Extended Data Fig. 1. Comparison of sol morphology in composites with and without PEIE.** SEM images of the composites containing **a,** 0 wt% PEIE and **b,** 1.5 wt% PEIE, with magnified views of the boxed regions. The sol areas are highlighted in blue. The PEIE concentrations are expressed relative to the Ecoflex mass. Scale bars, 20 µm (left) and 5 µm (right).

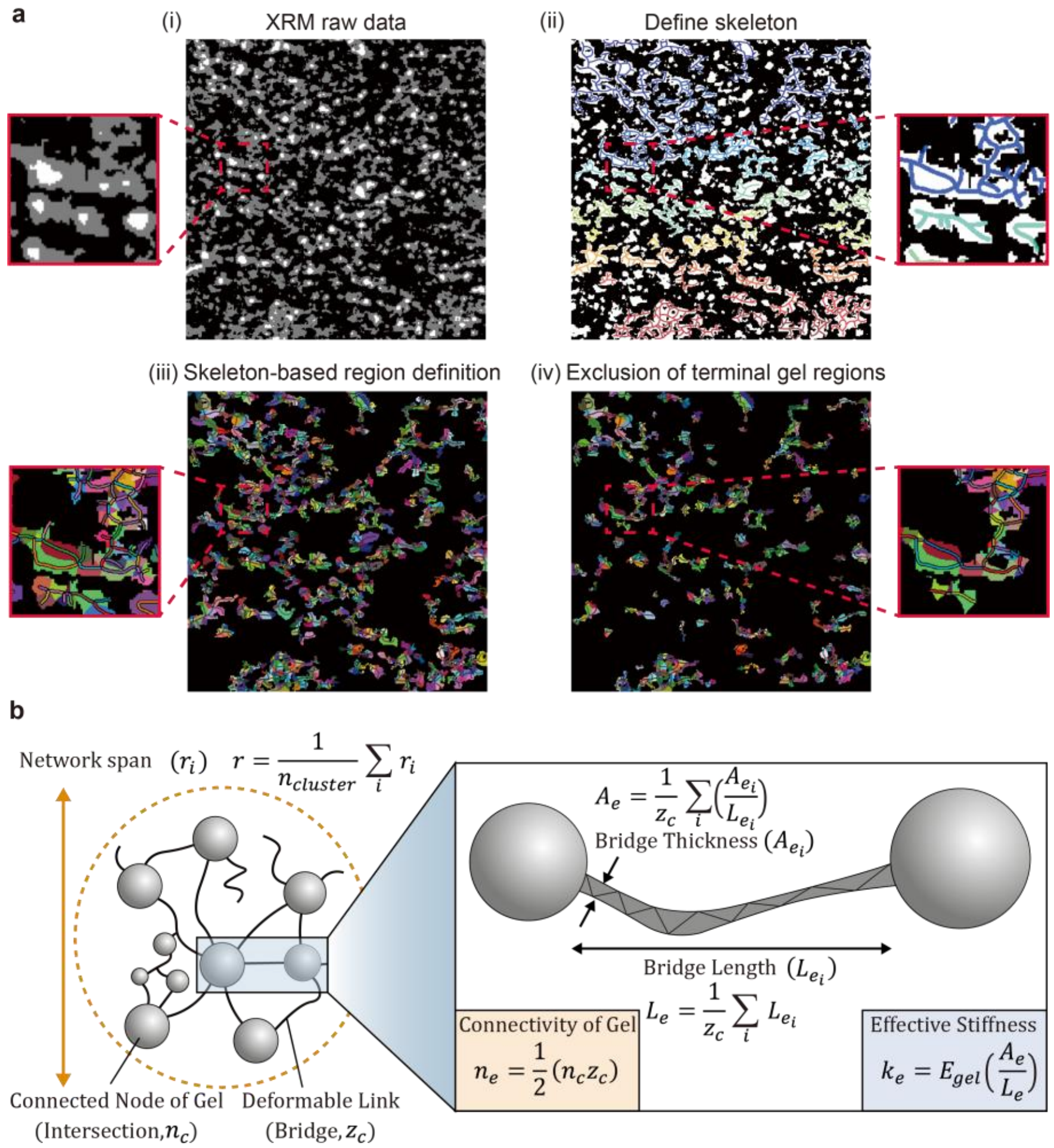

**Extended Data Fig. 2. Workflow for the XRM-based skeleton analysis of the gel topology and schematic of the load-bearing backbone used for the LCC-based modulus scaling analysis.**

**a,** (i) Representative XRM raw data with an enlarged view of the selected region. (ii) Skeleton definition of the binarized gel phase, with an enlarged view of the annotated structure. (iii) Skeleton-guided partitioning of the gel regions by assigning each pixel to the nearest skeleton branch. (iv) Final bridge map after exclusion of the terminal gel regions. This workflow was used to extract topological descriptors of the load-bearing gel skeleton. **b,** The gel morphology is represented as gel-rich domains connected by bridge-like ligaments. The mechanically relevant structure is defined as the largest connected component (LCC), from which the elastically active connectivity $n_e = \frac{1}{2}(n_c z_c)$, characteristic network span $\langle r^2 \rangle$, and average effective bridge stiffness $\bar{\kappa}$ are evaluated. In this framework, the modulus follows the scaling relation $G' \sim n_e \bar{\kappa} \langle r^2 \rangle$, emphasizing that global elasticity is governed by the integrity of the connected load-bearing backbone rather than by the total gel fraction alone.